\documentclass{PoS}

\usepackage{amsmath}
\usepackage{amssymb}

\newcommand{\Nt}{N_t}
\newcommand{\Nc}{N_c}

\newcommand{\lr}[1]{\left( #1 \right)}
\newcommand{\hmu}{{\hat{\mu}}}
\newcommand{\hnu}{{\hat{\nu}}}
\newcommand{\Lag}{\mathcal{L}}
\newcommand{\expval}[1]{\left\langle #1 \right\rangle}
\newcommand{\cbc}{\bar{\psi}\psi}

\title{Thermodynamics at Strong Coupling\\on Anisotropic Lattices}

\ShortTitle{Thermodynamics at Strong Coupling on Anisotropic Lattices}

\author{\speaker{Wolfgang Unger}\\
        Bielefeld University\\
        E-mail: \email{wunger@physik.uni-bielefeld.de}}

\author{Dennis Bollweg\\
        Bielefeld University\\
        E-mail: \email{dbollweg@physik.uni-bielefeld.de}}

\author{Marc Klegrewe\\
        Bielefeld University\\
        E-mail: \email{mklegrewe@physik.uni-bielefeld.de}}

\abstract{
Lattice QCD with staggered fermions at strong coupling has long been studied in a dual representation to circumvent
the finite baryon density sign problem. 
Monte Carlo simulations at finite temperature and density require anisotropic lattices.
Recent results that established the non-perturbative
functional dependence between the bare anisotropy $\gamma$ and the physical anisotropy $a_s/a_t$ in the
chiral limit are now extended to finite quark mass. We illustrate how the calibration of the anisotropy works and discuss the consequences of the anisotropy on thermodynamic observables. We also show first results on the energy density and pressure in the QCD phase diagram in the strong coupling regime.
}

\FullConference{The 36th Annual International Symposium on Lattice Field Theory - LATTICE2018\\
		22-28 July, 2018\\
		Michigan State University, East Lansing, Michigan, USA.}

\begin{document}

\newcommand{\mh}{\hat{m}}

\section{Introduction: Sign Problem}

Lattice QCD at finite baryon density suffers from the numerical sign problem: no direct simulations based on the fermion determinant (such as RHMC) are feasible, due to the fact that the fermion determinant becomes complex for $\mu_B>0$. Hence, with the well established methods, a possible critical endpoint is still out of reach. 
Methods based on a complexified parameter space (Complex Langevin, Lefschetz Thimbles) are promising, but not (yet) applicable to full QCD.
Since the sign problem is representation-dependent, the partition sum can also be rewritten in new degrees of freedom which are closer to the physical states. Then the sign problem can be milder or even be absent. Here, we will make use of so-called dual representations,. These have been proven useful in many models which have severe sign problems in the original formulation (e.~g.~\cite{Gattringer2015}).
For lattice QCD, a dual representation is well known in the strong coupling limit in terms of a monomer-dimer system \cite{Rossi1984,Karsch1989,Forcrand2010}.
In this limit $\beta=\frac{2\Nc}{g^2} \rightarrow 0$, it is possible to reverse the order of integration and integrate out all gauge fields $U_\mu(x)$ before the Grassman variables since link integration factorizes due to the absence of the plaquette contributions of the gauge action. 
The resulting color singlet degrees of freedom are mesons and baryons. This system has been studied since decades both via Monte Carlo and mean field and has proven to be a great laboratory for finite density QCD.
The advantage of the dual formulation of strong coupling LQCD is twofold: (1) the very mild sign problem (which is even absent in the continuous time limit) and (2) the applicability of Worm algorithms that enable fast simulations. This allows to study the full phase diagram in the $\mu_B$ - $T$ plane.

\section{The Dual Representation of Strong Coupling Lattice QCD}

The strong coupling partition function is obtained from the fermionic action of staggered fermions by an exact rewriting of the path integral by integrating out the gluons first.
Followed by Grassmann integration, it can be mapped on a discrete system of monomers $n_x \in \{0,\ldots \Nc\}$, dimers $k_b\in \{0,\ldots \Nc\}$, and world lines $\ell_b \in \{0,\pm 1\}$ \cite{Rossi1984,Karsch1989}:
\begin{equation}
Z_F(m_q,\mu)= \sum_{\{k,n,\ell\}}
\prod_{b=(x,\mu)}\frac{(\Nc-k_b)!}{\Nc!k_b!}
\prod_{x}\frac{\Nc!}{n_x!}(2am_q)^{n_x}
\prod_\ell \left(\frac{1}{\Nc!^{|\ell|}}\sigma(\ell) e^{3 \Nt r_l a_{\tau} \mu}\right)
\end{equation}
The Grassmann integration imposes the following constraint on the sum over configurations in the above partition sum:
\begin{align}
n_x+\sum_{\hat{\mu}=\pm\hat{0},\ldots \pm \hat{d}}
\lr{k_{\hat{\mu}}(x) + \frac{\Nc}{2} |\ell_\hmu(x)|} =\Nc,
\end{align}
which is a remnant of the gauge group and entails that mesonic degrees of freedom (monomers and dimers) do not touch baryon world lines, The latter form oriented self-avoiding loops $\ell$ of length $|\ell|$, and its sign $\sigma(\ell)\in\{-1,+1\}$ depends on loop geometry.

The caveat of this formulation is that the lattice is very coarse, and it requires $\beta>0$ to make the lattice finer.
In principle it is possible to include the effects of the gauge action by expanding it in terms of plaquette occupation numbers before integrating out the gauge links. This gives rise to a strong coupling expansion. Here, we do not include such corrections, but they have been addressed to next to leading order \cite{deForcrand2014, Unger2017}  and have also been presented at this conference in the contribution \cite{Gagliardi2018} ``Towards a Dual Representation of Lattice QCD'' by G.~Gagliardi. Alternative strategies 
with dual variables have been proposed in \cite{Gattringer2016,Borisenko2017}.
The leading order gauge correction $\mathcal{O}(\beta)$ to the  SC-LQCD phase diagram in the chiral limit has been first addressed via reweighting in $\beta$ from the ensemble at $\beta=0$ \cite{deForcrand2014}. Here it was found that although the nuclear liquid gas critical end point splits from the chiral tri-critical point, the first order line 
from the nuclear transition and the chiral transition did not split. 
An immediate question one can ask is: do the nuclear and chiral transition split at sufficiently large $\beta$? 
New simulations obtained by sampling plaquette contributions directly via world sheets did not indicate any splitting \cite{Unger2017}. 
We may have to consider the possibility that in the chiral limit, both transitions are on top even in the continuum limit. Hence it might be necessary to address simulations at finite quark mass for the splitting to be sizeable. This motivates the study presented here.

\section{Thermodynamics of Strong Coupling Lattice QCD}

In order to vary the temperature in the strong coupling limit, where the lattice spacing $a(\beta)$ cannot be modified at fixed $\beta=0$,
we need to introduce the bare anisotropy $\gamma$ in the Dirac couplings.
This is in particular necessary since $aT=1/\Nt$ is discrete (with $\Nt$ even): it turns out that the chiral transition temperature is about $aT_c\simeq 1.5$, hence we cannot address the phase transition on isotropic lattices. The bare anisotropy will change the temporal lattice spacing $a_t$ continuously at fixed $a_s\equiv a$.
: 
\begin{align}
\Lag_{\rm F}(\mh,a_t\mu,\gamma)&= \sum_{x}\left\{\sum_{\nu} \gamma^{\delta_{\nu 0}}\eta_\nu(x)\left(e^{a_t\mu\delta_{\nu 0}} \bar{\psi}_x U_\nu(x) \psi_{x+\hnu} - e^{-a_t\mu\delta_{\nu 0}}  \bar{\psi}_{x+\hnu} U_\nu^\dagger(x) \psi_x \right)+2\mh\bar{\psi}_x\psi_x\right\}\nonumber\\
Z_F(\mh,a_t\mu,\gamma)&= \sum\limits_{\{k,n,\ell\}}\prod\limits_{b=(x,\nu)}\frac{(\Nc-k_b)!}{\Nc!k_b!}
\gamma^{2 k_b\delta_{\nu 0}}\prod\limits_{x}\frac{\Nc!}{n_x!}(2\mh)^{n_x} \prod\limits_\ell w(\ell,a_t\mu)
\label{SC}
\end{align}
The anisotropy $\frac{a_s}{a_{t}}\equiv \xi(\gamma)$ is a non-perturbative function of the bare anisotropy $\gamma$
which allows to define the temperature  $aT=\frac{\xi(\gamma)}{\Nt}$.
At weak coupling one expects $\xi(\gamma)=\gamma$, however, at strong coupling, where the degrees of freedom are not quarks but hadrons, this is not the case. Mean field theory at strong coupling implies $\xi(\gamma)=\gamma^{2}$, since the square of the critical bare coupling is proportional to $\Nt$: $\gamma_c^2=\Nt \frac{(d-1)(\Nc+1)(\Nc+2)}{6(\Nc+3)}$ \cite{Bilic1992a}. However, modifications are expected beyond mean field, hence we need to determine the precise correspondence between $\xi\equiv a_s/a_t$ and $\gamma$.

Consider the SU(3) partition function Eq.~(\ref{SC}), in terms of the extensive quantities: $N_M=\sum\limits_x n_x$ the total monomer number, $N_q=2N_{Dt}+3N_{Bt}$ 
(with $N_{Dt}=\sum\limits_x k_{x,0}$ and $N_{Bt}=\sum\limits_{x}|b_{x,0}|$ 
the total number of temporal dimer and temporal baryon segments), and $\Omega$ the total winding number of all baryon world lines.
The dimensionless thermodynamic observables in terms of these dual variables are:
\begin{align}
\text{baryon density:} &&  a_s^3\rho_B&=\left.a_s^3\frac{T}{V}\frac{\partial \log Z}{\partial \mu_B}\right|_{V,T}=\frac{\expval{\Omega}}{N_\sigma^3}=\expval{\omega}
\label{Obs1}
\\
\text{energy density:} &&  a_s^3a_t\epsilon&=\mu_B\rho_B-\left.\frac{a^3a_t}{V}\frac{\partial \log Z}{\partial T^{-1}}\right|_{V,\mu_B}=\frac{\xi}{\gamma}\frac{d\gamma}{d\xi} \expval{n_q}-\expval{n_M}\\
\text{pressure:}&& a_s^3a_tp&=-\left.a_s^3a_tT\frac{\partial \log Z}{\partial V}\right|_{T,\mu_B}=\frac{\xi}{3\gamma}\frac{d\gamma}{d\xi}\expval{n_q}
\end{align}
\newpage
\vphantom{.}
\vspace{-15mm}
\begin{align}
\text{chiral condensate:}&&  a_s^4 \expval{\cbc}&=a_s^4 \frac{\expval{N_M}}{N_\sigma^3 \Nt a_s^4a_t}=\frac{\xi}{\mh}\expval{n_M}\label{Obs2}
\\
\text{interaction measure:}&& \epsilon-3p&=-\frac{\expval{n_M}}{a_s^3a_t}=-m_q\expval{\cbc  }
\end{align}

Clearly, most of these observables explicitly depend on $\xi(\gamma)$ and its derivative. They have been measured in the full $\mu_B$ - $T$ plane after having determined $\xi(\gamma)$ non-perturbatively.

\section{Anisotropy Calibration and Results}
 
The determination of $\xi(\gamma)$ in the chiral limit has already been addressed in \cite{deForcrand2016,deForcrand2017}. The non-perturbative result deviates from 
the mean field assignment $\xi_{\rm mf}(\gamma)=\gamma^2$ considerably:
\begin{align}
\xi(\gamma)&\approx \kappa\gamma^2+\frac{\gamma^2}{1+\lambda\gamma^4}\qquad \text{with} \qquad \kappa=0.781(1)
\end{align}
As an application, the dependence of observables on the anisotropy was studied: the pion decay constant, the  chiral condensate  and the baryon mass. With this result, it is also possible to define unambiguously the continuous time limit 
$a_t\rightarrow 0$ via $\Nt\rightarrow \infty$ and $\gamma\rightarrow \infty$  at fixed $aT$,
which is further elaborated in \cite{Unger2012} and in a contribution to this conference ``Temporal Correlators in the Continuous Time Limit of Lattice QCD''
my M.~Klegrewe \cite{Klegrewe2018}. The anisotropy calibration can also directly be performed in the continuous time limit.

In this proceedings, we want to extend these results to finite quark, i.~e.~we address the anisotropy calibration and its difficulties for $m_q>0$. 
In order to determine $\xi(\gamma)$ the idea is to consider the following current that is implied by the Grassmann constraint \cite{Chandrasekharan2003}: 
 \begin{align}
 j_\mu(x)&=\sigma(x)\left(k_\mu(x)-\frac{\Nc}{2}|b_{x,\mu}|-\frac{\Nc}{2d}\right)\quad \rightarrow \quad \sum\limits_{\pm\hmu}(j_\mu(x)-j_\mu(x-\hmu))=-\sigma(x)n(x)
 \end{align}
 In the chiral limit, where $n(x)=0$, the current $j_\mu(x)$ is locally conserved.
 The conserved charge $Q_\mu=\sum\limits_{x\perp \mu}j_\mu(x)$ has $\expval{Q_\mu}=0$, but non-zero variance: $\expval{Q_\mu^2}\neq0$.
 The calibration of $\xi(\gamma)$ is then obtained via a renormalization condition on demanding a physically isotropic box:
\begin{align}
a_t N_t&=a_s N_s \quad \Leftrightarrow \quad \expval{Q_t^2}(\gamma_0)\stackrel{!}{=} \expval{Q_s^2}(\gamma_0),
 \qquad \frac{a_s}{a_t}=\frac{N_t}{N_s}=\xi(\gamma_0).
 \end{align}

 \begin{figure}[h!]
 \includegraphics[width=\textwidth]{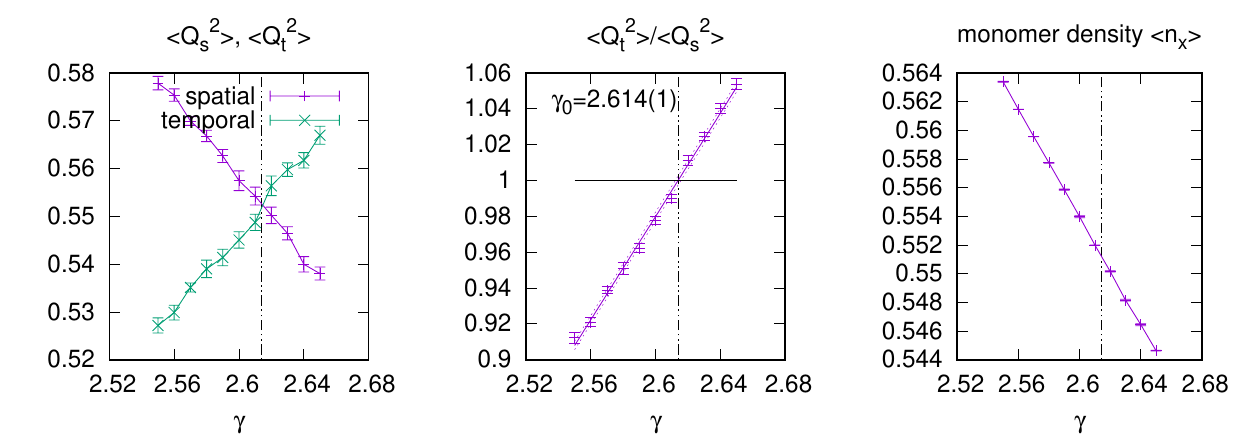}  
 \caption{
Anisotropy calibration at bare mass $\hat{m}=0.1$, $\xi=5$ on a lattice $10^3\times 50$. 
\emph{Left:} Spatial and temporal charge fluctuations. 
\emph{Center:} Ratio, from which $\gamma_0$ is obtained.
\emph{Right:} Monomer density around $\gamma_0$. 
}
\label{Cal1}
\end{figure}

\begin{figure}
\centerline{
\includegraphics[width=0.43\textwidth]{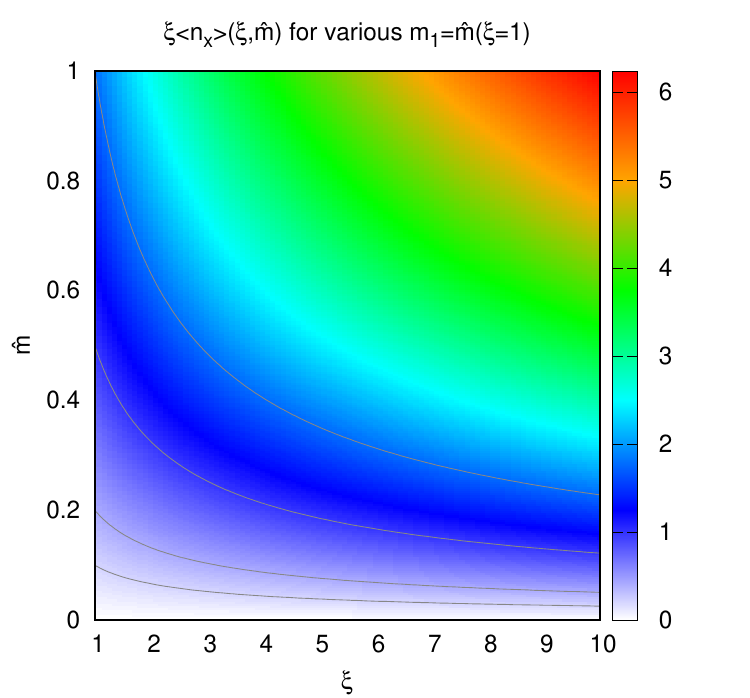}\quad
\includegraphics[width=0.43\textwidth]{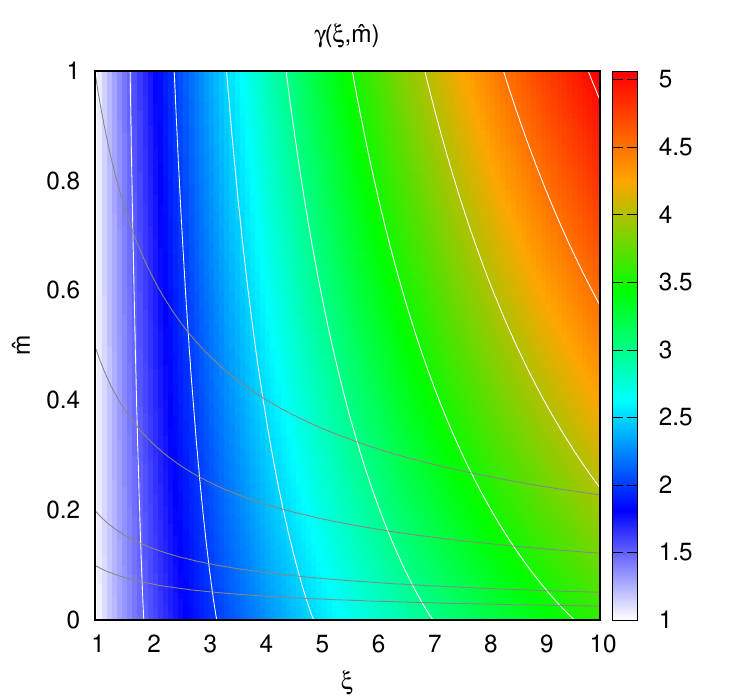}
}
\caption{Anisotropy calibration at finite quark mass: 
\emph{Left:} Lines of constant physics: fixed $m_q\cbc$, which defines $\mh(\xi)$ with $\mh_1=\mh(\xi=1)=am_q$.
\emph{Right:}
bare anisotropy $\gamma$ as a function of $\xi$ and $\mh$.}
\label{Cal2}
\end{figure}

\begin{figure}
\centerline{\includegraphics[width=0.8\textwidth]{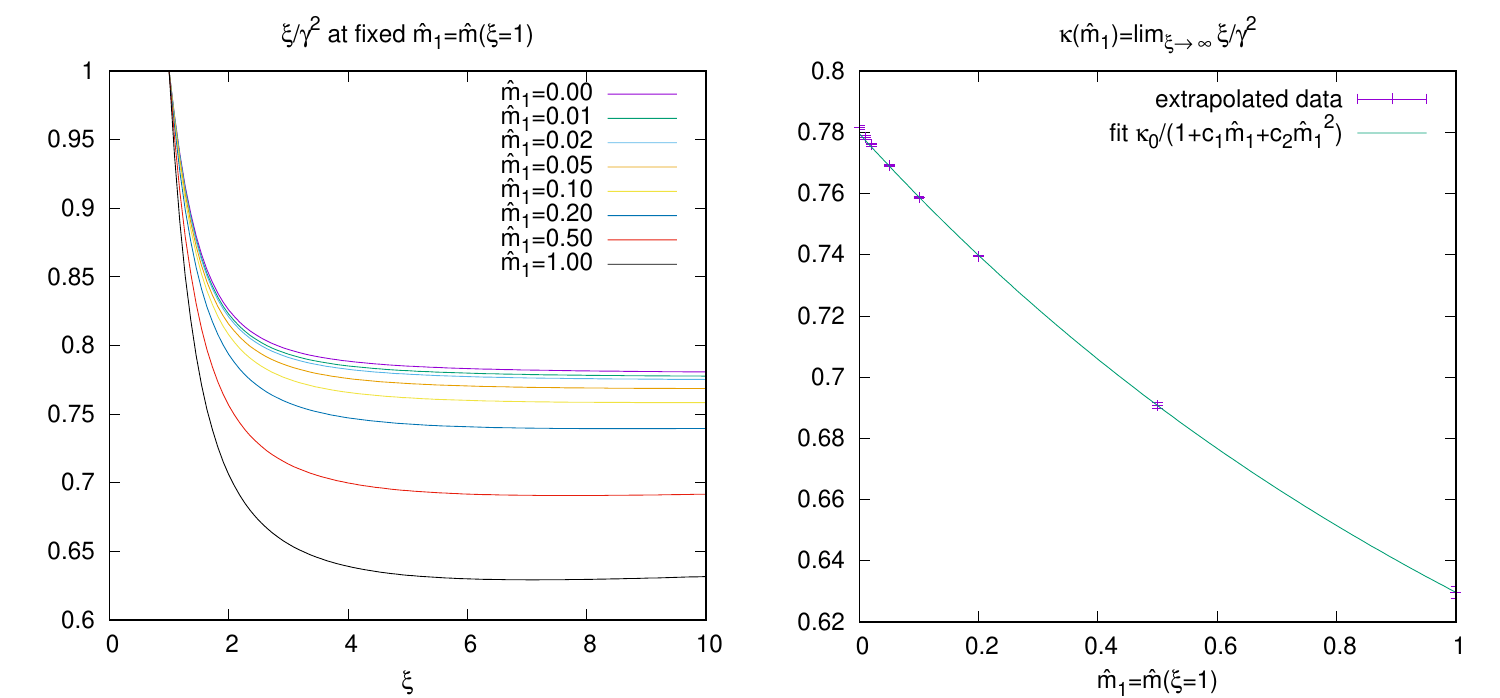}}
\caption{Example on how to obtain the continuous time limit at finite quark mass. \emph{Left:} Correction factor $\xi(\gamma)/\gamma^2$ at fixed $\mh_1$. \emph{Right:} Dependence of the correction factor $\kappa$  in the continuous time limit on $m_q$.}
\label{CTLimit}
\end{figure}

To extend this method to finite quark mass, 
there is yet a difficulty:
 $j_\mu(x)$ is no longer a conserved current, i.e.~on a given configuration, $Q_t(t_1) \neq Q_t(t_2)$ (and $Q_z(z_1) \neq Q_t(z_2)$).
This is expected because the monomers are sources of a pion current $-m_q\bar{\psi}\gamma_5\psi$.
However, due to the even/odd decomposition for staggered fermions, there are as many monomers on even as on odd sites. 
Hence, when averaging over parallel hypersurfaces, 
\begin{align}
Q_t&=\frac{1}{N_t}\sum_t(Q_t(t)),& Q_z&=\frac{1}{N_s}\sum_z(Q_z(z)), 
\end{align}
 the monomers of opposite parity $\sigma(x)$ cancel each other, 
such that the total charge and its fluctuations can still be used for anisotropy calibration. 
 Again, we demand the fluctuations to be equal, $\expval{Q_t^2}(\gamma_0)\stackrel{!}{=} \expval{Q_s^2}(\gamma_0)$.
 However, at finite dimensionless bare quark mass $\mh$, we need to keep the physics constant e.~g.~$M_\pi L =\text{const}$
 or $[m_q\expval{\cbc}]_L =\text{const}$.
Hence  we need to determine $\mh(\xi)$ as well (see also \cite{Levkova2006}).
We implement the second condition:
\begin{align}
a^4 m_q \expval{\cbc}&= a^3 a_t\xi \mh(\xi)\expval{\cbc} = \xi\expval{n_x}={\rm const},
\end{align}
which is related to the monomer density. Note that it is not possible to identify $\mh$ with either $am_q$ nor $a_tm_q$ as $\mh$ depends on 
$\xi$ ($\mh$ is the bare mass in Eq.~(\ref{SC})).
In Fig.~\ref{Cal1} an example of the anisotropy calibration is shown. Fig.~\ref{Cal2} shows the final result obtained by scanning through various bare quark masses $\mh\in[0,1]$ and lattices $10^3\times N_t$ with aspect ratio $\frac{N_t}{N_s}=\xi\in \{1,2,3,4,5,6,8,10\}$.
With the calibration results, the continuous time limit $N_t\rightarrow \infty$ is well defined also for finite $m_q$ with $m_q/T$ fixed. The extrapolation towards continuous time is shown in Fig.~\ref{CTLimit}.
The non-perturbative correction factor turns
out to have a simple quark mass dependence, such that the temperature and chemical potential are uniquely specified
and have a well defined continuous time limit also at finite quark mass $\mh_1=am_q$:
\begin{align}
\kappa(\mh_1)&=\frac{\kappa_0}{1+c_1 \mh_1+c_2 \mh_1^2} & aT&=\kappa(\mh_1)[aT]_{\rm mf}& a\mu_B&=\kappa(\mh_1)[a\mu_B]_{\rm mf}
\end{align}

\begin{figure}
\centerline{
\vspace{-4mm}
\includegraphics[width=0.43\textwidth]{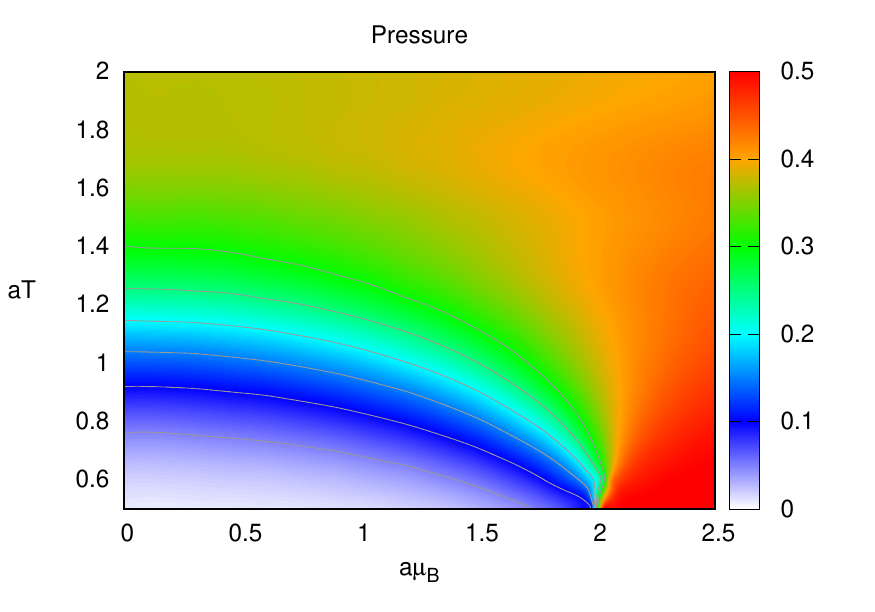}
\includegraphics[width=0.43\textwidth]{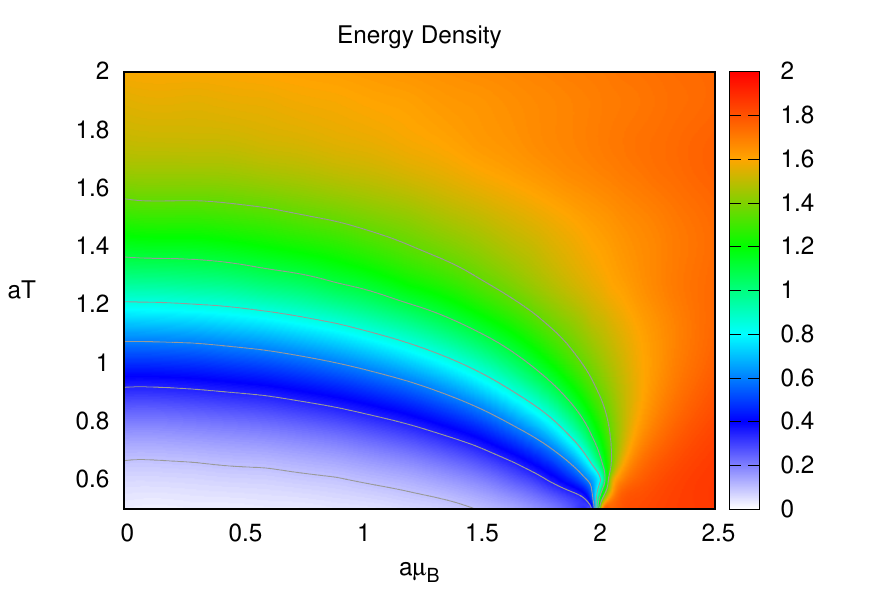}}
\centerline{
\includegraphics[width=0.43\textwidth]{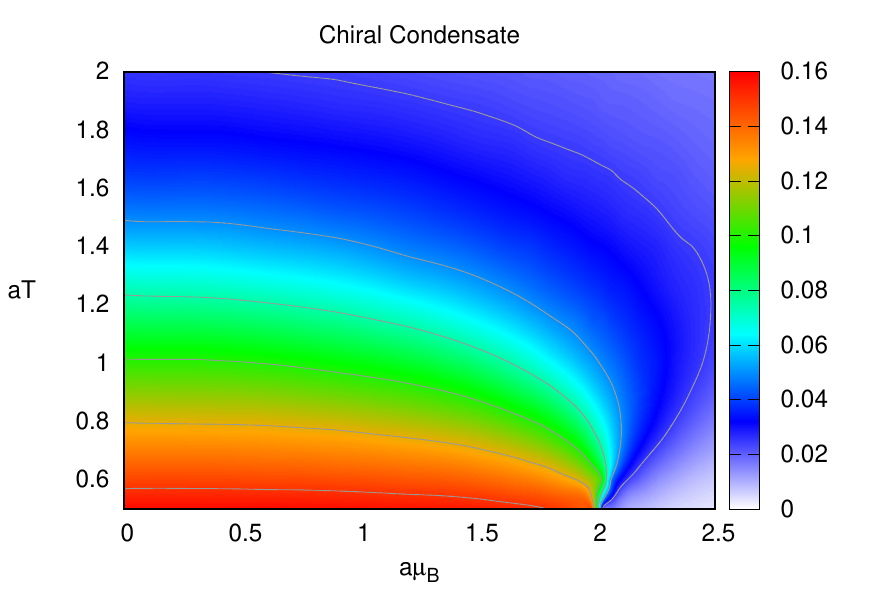}
\includegraphics[width=0.43\textwidth]{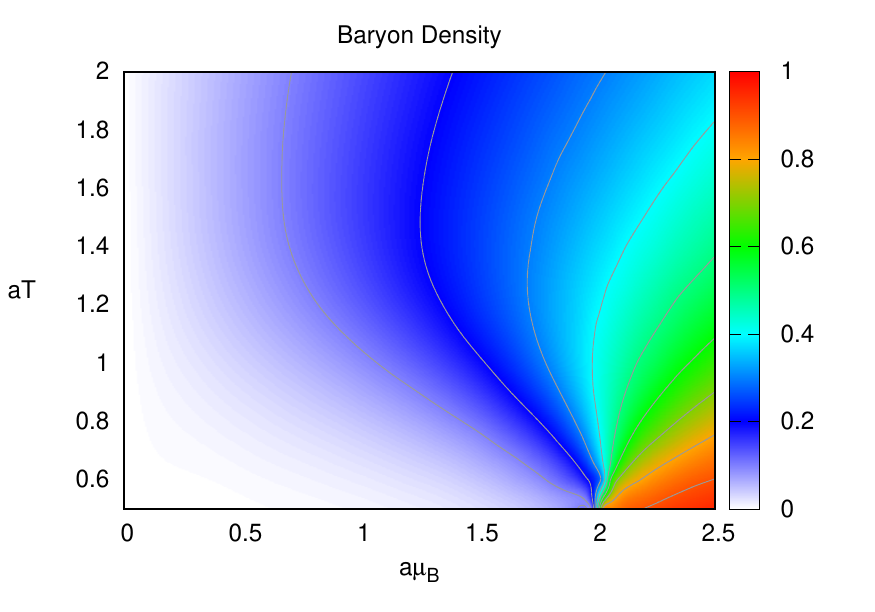}}\vspace{-4mm}
\caption{Various thermodynamic observables in the $\mu_B$ - $T$ plane at quark mass $am_q=0.1$. The imprint of the chiral critical endpoint at around 
$(aT,a\mu_B)\simeq(0.6,2.0)$ can clearly be seen in all observables.}
\label{Obs}
\end{figure}

By fixing the temperature $aT$ on a lattice specified by $\Nt$, and the isotropic bare quark mass $\mh_1=am_q$, one can determine via 
$\xi=aTN_t$ the corresponding $\mh(\xi)$ and $\gamma_0(\xi,\mh(\xi))$ for the Monte Carlo simulations.
With this it is possible to measure various thermodynamic observables at fixed mass in the $\mu_B$-$T$ plane. 
Results on the phase boundary  at $m_q>0$ have already been addressed in \cite{Kim2016}, but here the mass-independent mean field definitions $aT$ and $a\mu_B$ were used. The new results for the thermodynamic observables Eqs.~(\ref{Obs1})-(\ref{Obs2}), 
are shown in Fig.~(\ref{Obs}) for $am_q=0.1$.

\definecolor{ar}{rgb}{0.0, 0.7, 0.0}

\section{Conclusions}
 
We have shown how to extend the anisotropy calibration to finite quark mass to obtain    the bare anisotropy $\gamma_0(\xi,\mh(\xi))$ given 
$\xi=\frac{a_s}{a_t}$ corresponds to a physically isotropic lattice: $a_sN_s=a_t\Nt$. 
Here, the difficulty was addressed that $\gamma_0$ now also depends on $\mh$, which requires an additional condition that keeps the physics constant and yields $\mh=\mh(\xi)$.
This allows us to define the temperature/chemical potenital and measure thermodynamic observables such as energy and pressure unambiguously.
Simulations in the continuous time limit $\xi\rightarrow \infty$ confirm the extrapolated results (see also the contribution to this conference \cite{Klegrewe2018}).
In the future, we want to address the anisotropy calibration also for $\beta>0$: Here, the non-perturbative determination of $a_s/a_t\equiv \xi(\gamma,\mh,\beta)$ now also involves $\beta$, and it might be necessary to introduce an additional bare anisotropy $\beta_s/\beta_t$ as well.

\section{Acknowledgment}

We thank Philippe de Forcrand and H\'elvio Vairinhos for useful discussions.
Numerical simulations were performed on the OCuLUS cluster at PC$^2$ (Universit\"at Paderborn). We acknowledge support by the Deutsche Forschungsgemeinschaft (DFG) through the
Emmy Noether Program under Grant No.~UN 370/1 and through the Grant No.~CRC-TR 211
``Strong-interaction matter under extreme conditions''. 
    
  \bibliography{Unger2} 

\providecommand{\href}[2]{#2}\begingroup\raggedright\begin{thebibliography}{10}

\bibitem{Gattringer2015}
C.~Gattringer, T.~Kloiber and V.~Sazonov,
  ~\href{https://doi.org/10.1016/j.nuclphysb.2015.06.017}{\emph{Nucl. Phys.}
  {\bfseries B897} (2015) 732}
  [\href{https://arxiv.org/abs/1502.05479}{{\ttfamily 1502.05479}}].

\bibitem{Rossi1984}
P.~Rossi and U.~Wolff,
  ~\href{https://doi.org/10.1016/0550-3213(84)90589-3}{\emph{Nucl. Phys.}
  {\bfseries B248} (1984) 105}.

\bibitem{Karsch1989}
F.~Karsch and K.~H. Mutter,
  ~\href{https://doi.org/10.1016/0550-3213(89)90396-9}{\emph{Nucl. Phys.}
  {\bfseries B313} (1989) 541}.

\bibitem{Forcrand2010}
P.~de~Forcrand and M.~Fromm,
  ~\href{https://doi.org/10.1103/PhysRevLett.104.112005}{\emph{Phys. Rev.
  Lett.} {\bfseries 104} (2010) 112005}
  [\href{https://arxiv.org/abs/0907.1915}{{\ttfamily 0907.1915}}].

\bibitem{deForcrand2014}
P.~de~Forcrand, J.~Langelage, O.~Philipsen and W.~Unger,
  ~\href{https://doi.org/10.1103/PhysRevLett.113.152002}{\emph{Phys. Rev.
  Lett.} {\bfseries 113} (2014) 152002}
  [\href{https://arxiv.org/abs/1406.4397}{{\ttfamily 1406.4397}}].

\bibitem{Unger2017}
G.~Gagliardi, J.~Kim and W.~Unger,
  ~\href{https://doi.org/10.1051/epjconf/201817507047}{\emph{EPJ Web Conf.}
  {\bfseries 175} (2018) 07047}
  [\href{https://arxiv.org/abs/1710.07564}{{\ttfamily 1710.07564}}].

\bibitem{Gagliardi2018}
G.~Gagliardi and W.~Unger, ~{\emph{PoS} {\bfseries LATTICE2018} (2018) 224}
  [\href{https://arxiv.org/abs/1811.02817}{{\ttfamily 1811.02817}}].

\bibitem{Gattringer2016}
C.~Gattringer and C.~Marchis,
  ~\href{https://doi.org/10.1016/j.nuclphysb.2017.01.025}{\emph{Nucl. Phys.}
  {\bfseries B916} (2017) 627}
  [\href{https://arxiv.org/abs/1609.00124}{{\ttfamily 1609.00124}}].

\bibitem{Borisenko2017}
O.~Borisenko, V.~Chelnokov and S.~Voloshyn,
  ~\href{https://doi.org/10.1051/epjconf/201817511021}{\emph{EPJ Web Conf.}
  {\bfseries 175} (2018) 11021}
  [\href{https://arxiv.org/abs/1712.03064}{{\ttfamily 1712.03064}}].

\bibitem{Bilic1992a}
N.~Bilic, F.~Karsch and K.~Redlich,
  ~\href{https://doi.org/10.1103/PhysRevD.45.3228}{\emph{Phys. Rev.} {\bfseries
  D45} (1992) 3228}.

\bibitem{deForcrand2016}
P.~de~Forcrand, P.~Romatschke, W.~Unger and H.~Vairinhos,
  ~\href{https://doi.org/10.22323/1.256.0086}{\emph{PoS} {\bfseries
  LATTICE2016} (2017) 086} [\href{https://arxiv.org/abs/1701.08324}{{\ttfamily
  1701.08324}}].

\bibitem{deForcrand2017}
P.~de~Forcrand, W.~Unger and H.~Vairinhos,
  ~\href{https://doi.org/10.1103/PhysRevD.97.034512}{\emph{Phys. Rev.}
  {\bfseries D97} (2018) 034512}
  [\href{https://arxiv.org/abs/1710.00611}{{\ttfamily 1710.00611}}].

\bibitem{Unger2012}
W.~Unger and P.~de~Forcrand,
  ~\href{https://doi.org/10.22323/1.164.0194}{\emph{PoS} {\bfseries
  LATTICE2012} (2012) 194} [\href{https://arxiv.org/abs/1211.7322}{{\ttfamily
  1211.7322}}].

\bibitem{Klegrewe2018}
M.~Klegrewe and W.~Unger, ~{\emph{PoS} {\bfseries LATTICE2018} (2018) 182}
  [\href{https://arxiv.org/abs/1811.01614}{{\ttfamily 1811.01614}}].

\bibitem{Chandrasekharan2003}
S.~Chandrasekharan and F.-J. Jiang,
  ~\href{https://doi.org/10.1103/PhysRevD.68.091501}{\emph{Phys. Rev.}
  {\bfseries D68} (2003) 091501}
  [\href{https://arxiv.org/abs/hep-lat/0309025}{{\ttfamily hep-lat/0309025}}].

\bibitem{Levkova2006}
L.~Levkova, T.~Manke and R.~Mawhinney,
  ~\href{https://doi.org/10.1103/PhysRevD.73.074504}{\emph{Phys. Rev.}
  {\bfseries D73} (2006) 074504}
  [\href{https://arxiv.org/abs/hep-lat/0603031}{{\ttfamily hep-lat/0603031}}].

\bibitem{Kim2016}
J.~Kim and W.~Unger, ~\href{https://doi.org/10.22323/1.256.0035}{\emph{PoS}
  {\bfseries LATTICE2016} (2016) 035}
  [\href{https://arxiv.org/abs/1611.09120}{{\ttfamily 1611.09120}}].

\end{thebibliography}\endgroup
  \bibliographystyle{JHEP}

\end{document}